\documentclass[grl]{agu2001}
 \usepackage{graphicx}

\authorrunninghead{LEVERMANN}

\titlerunninghead{SOLUTION OF A MODEL FOR THE OCEANIC PYCNOCLINE DEPTH}

\authoraddr{Anders Levermann,
Climate System Department, Potsdam Institute for Climate Impact Research, 
Telegraphenberg A25, 14473 Potsdam, Germany.
(Anders.Levermann@pik-potsdam.de)}

\begin{document}

\title{Solution of a Model for the Oceanic Pycnocline Depth: Scaling of 
Overturning Strength and Meridional Pressure Difference}

\author{A. Levermann and A. Griesel}
\affil{Climate System Department, Potsdam Institute for Climate Impact Research, Potsdam, Germany}

\begin{abstract}
We present an analysis of the model by 
{\it Gnanadesikan} [1999] 
for the pycnocline depth in the ocean. An 
analytic solution for the overturning strength as a function of the 
meridional pressure difference is derived and used to
discuss their mutual scaling.
We show that scaling occurs only in two unphysical regimes of the model.
In the absence of the Southern Ocean (SO) processes, i.e. for a northern overturning 
cell, the volume transport is proportional to the square root of the 
pressure difference.
Linear scaling is seen when the overturning is restricted entirely to the 
SO, i.e. when no northern downwelling exists.
For comparison, we present simulations with the coupled climate model 
CLIMBER-3$\alpha$ which show linear scaling over a large regime of pressure 
differences in the North Atlantic (NA). We conclude that the pycnocline 
model is not able to reproduce the linear scaling between its two central 
variables, pressure and volume transport.
\end{abstract}

\begin{article}

\section{Introduction}
The meridional overturning circulation in the Atlantic is a central challenge 
to our understanding of global climate dynamics. {\it Gnanadesikan} [1999] ({\it G99} hereafter)
presented a model for 
the deep meridional circulation in terms of the pycnocline depth (PD). 
This idealised model has been under intense investigation as a possible paradigm for the 
meridional overturning circulation 
[{\it Gnanadesikan and Hallberg}, 2000; {\it Saenko and Weaver}, 2002; 
{\it Gnanadesikan et al.}, 2002].
Furthermore it has been used to investigate the qualitative importance of 
different physical feedbacks on the oceanic circulation 
[{\it Klinger et al.}, 2003, {\it Gnanadesikan et al.}, 2003; {\it Kamenkovich and Sarachik}, 2004]. 
A qualitative feature of the deep meridional overturning circulation is the  
scaling relation between the volume transport and the meridional density difference 
in the Atlantic [{\it Bryan} 1987]. Picking up Bryan's 
scaling arguments but assuming a constant PD in the Atlantic 
{\it Rahmstorf} [1996] proposed a linear relation 
which he demonstrated in the oceanic general circulation model (GCM) MOM-2. 
{\it Park} [1999] and {\it Scott et al.} [1999] derived the same scaling in a 
Stommel-type box model.
GCM simulations of the ocean suggest that this linear relation 
carries over from the density to the pressure difference 
[{\it Hughes and Weaver}, 1994; {\it Thorpe et al.}, 2001]. The linear scaling 
relation between pressure difference and maximum overturning strength 
has since been demonstrated to be a robust feature in oceanic GCM 
simulations ({\it Griesel}, pers. comm., 2004).
In sec.~4 we present simulations with the coupled climate model 
CLIMBER-3$\alpha$ further supporting these findings.

The {\it G99} model contains four physical processes which influence the 
PD in the ocean. The balance of the pressure gradient in the North 
Atlantic and 
the frictional forces within the boundary currents leads to an equation 
for the northward volume transport
\begin{equation}
T_n 
= \frac{C D}{\beta L_y^{(n)}} \cdot \frac{\Delta p}{\rho}  
= \frac{C g \Delta \rho}{\rho \beta L_y^{(n)}} \cdot D^2
\equiv \gamma_n g \Delta \rho \cdot D^2
\label{equ_tn}
\end{equation}
The pressure gradient is parameterised through the density difference in the 
NA $\Delta \rho$, the north-south distance $L_y^{(n)}$ over which 
the gradient occurs and the PD $D$. 
\begin{equation}
\Delta p = g D \Delta \rho
\label{equ_dp}
\end{equation}
The constant $\gamma_n$ combines $L_y^{(n)}$ with $\beta$, $\rho$ and $C$ 
(the meridional derivative of the Coriolis parameter $f$, the density and 
a proportionality constant of order one). $g$ is the gravity constant.
The quadratic dependence 
on $D$ occurs due to the vertical integration in order to obtain a volume 
transport.
In the SO the model includes the Drake passage effect through 
a wind-driven upwelling
which does not explicitly depend on the PD 
$T_s^{(e)} = (L_x \tau)/(\rho f) \equiv 2 \gamma_e$. $\tau$ and $L_x$ are the wind 
stress in the SO and the circumference around Earth at 
the latitude of Drake Passage. 
Additionally {\it G99} includes an eddy induced return flow
\begin{equation}
T_s^{(gm)} =  L_x v_{ed} \cdot D \equiv \gamma_{gm} \cdot D
\label{equ_ts_gm}
\end{equation}
where $v_{ed}$ is the transport velocity which {\it G99} parameterised 
following {\it Gent and McWilliams} [1990] while we focus here on its 
dependence on the PD. 
The fourth term in the model is associated with low-latitudinal 
upwelling described by an advection-diffusion balance 
$w \partial_z\rho = K_v \partial_{zz} \rho $ in the tropics which yields
\begin{equation}
T_u = \frac{K_v A_u}{D} \equiv \frac{\gamma_u}{D}
\label{equ_tu}
\end{equation}
where $K_v$ and $A_u$ are the diapycnal diffusivity and the horizontal area 
of upwelling, respectively. All non-negative constants $\gamma_x$ have been introduced 
for convenience. Note that the underlying assumption of the model 
is that these four process can be described using the same value $D$ for the 
PD throughout the Atlantic. Equ.~(\ref{equ_tn}) requires 
futhermore that the vertical extension of the northward volume flow is also 
given by $D$. Accepting these assumptions, the conservation of volume then 
results in the governing equation of the model
\begin{equation}
0 = \gamma_n g \Delta \rho \cdot D^3
+ \gamma_{gm} \cdot D^2 - 2 \gamma_e \cdot D
- \gamma_u \label{equ_governing}
\end{equation}
It can be shown that 
for all parameter settings the model has 
at most one solution with non-negative PD.
In sec.~\ref{sec_solution} we give this solution analytically in terms of the volume 
transport $T_n$ as a function of the pressure difference $\Delta p$ and discuss, in 
sec.~\ref{sec_scaling}, its scaling with $\Delta p$. In sec.~4 
we compare the results with simulations with the coupled climate model CLIMBER-3$\alpha$.

\section{Solution for the Volume Transport $T_n$} \label{sec_solution}
\begin{figure}%[h!]
\begin{center}
\includegraphics[width=18pc]{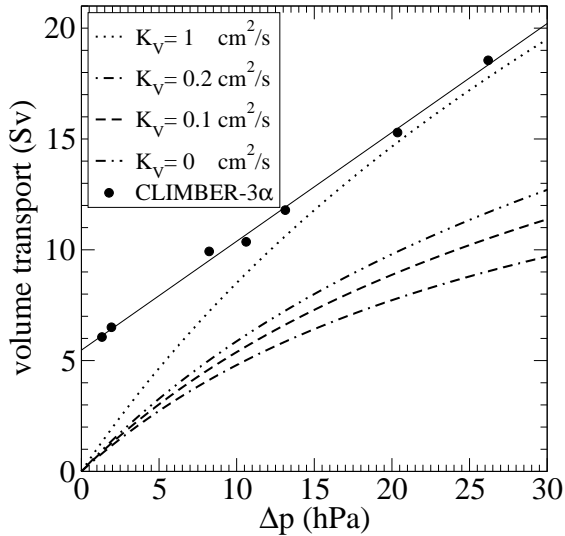}
\caption{\label{fig_analytic_solution} The analytic solution of the conceptual model for 
the volume transport 
$T_n$ as a function of the meridional pressure difference $\Delta p$ for different values 
of the diapycnal diffusivity $K_v$. The dots represent 
simulations with the coupled model CLIMBER-3$\alpha$ which contains 
an oceanic GCM. The simulations exhibit a linear scaling in 
contrast to the conceptual model.}
\end{center}
\end{figure}

In order to obtain an analytic solution of the model we rewrite 
equ.~(\ref{equ_tn}) to get an expression of the PD as a function 
of volume transport $T_n$ and pressure difference $\Delta p$
\begin{equation}
D
= T_n/(\gamma_n \Delta p) 
\label{equ_D_tn_dp}
\end{equation}
In the most interesting case of non-zero volume transport, $T_n \not= 0$, we 
can insert the equality~(\ref{equ_D_tn_dp}) into the volume conservation 
equ.~(\ref{equ_governing}) to get
\begin{equation}
0 = T_n + \frac{\gamma_{gm}}{\gamma_n} \frac{T_n}{\Delta p}  
- 2 \gamma_e - \gamma_u \gamma_n \frac{\Delta p}{T_n} 
\label{equ_transport}
\end{equation}
Multiplying by $T_n$ yields a quadratic equation in $T_n$ with two solutions of 
which only one is non-negative
\begin{equation}
T_n = \frac{\gamma_n \Delta p}{\gamma_n \Delta p + \gamma_{gm}} 
\left( \gamma_e + \sqrt{\gamma_e^2 + \gamma_u \left( \gamma_n \Delta p + \gamma_{gm}\right)}\right)
\label{equ_solution}
\end{equation}
Note that despite the fact that the governing equ.~(\ref{equ_governing}) is 
cubic in $D$, the model does have at most one physical solution given by 
equ.~(\ref{equ_solution}).
The model does therefore not bear the 
possibility of multiple stable modes of the deep meridional overturning circulation 
as suggested by simulations with climate models of different complexity
[{\it Stommel}, 1961; {\it Manabe and Stouffer}, 1988; {\it Rahmstorf}, 1995, 1996; 
{\it Ganopolski et.al.}, 2001; {\it Prange et al.}, 2003].
This is to be expected given that the model does not include a salt-advection feedback 
as proposed by 
{\it Stommel} [1961]. 
Fig.~\ref{fig_analytic_solution} shows the solution for different diapycnal mixing 
coefficients $K_v$. The results were obtained using the numerical values given by 
{\it G99}. 
Note that the solution (\ref{equ_solution}) depends continuously on the diapycnal 
mixing coefficient $K_v \propto \gamma_u$. No change in the quality of the 
solution~(\ref{equ_solution}) occurs in the absence of the low-latitudinal upwelling,
where 
\begin{equation}
T_n^{(K_v=0)} = \frac{2 \gamma_e \gamma_n \Delta p}{\gamma_n \Delta p + \gamma_{gm}} 
= \frac{\gamma_n \Delta p}{\gamma_n \Delta p + \gamma_{gm}} \cdot T_s^{(e)}
\label{equ_nokv_solution}
\end{equation}
In contrast to the behaviour for vanishing $K_v$, the elimination of the SO 
processes changes the quality of the solution as can be seen from equ.~(\ref{equ_solution}) 
and will be discussed in the next section.

%-figure:- analytic_solution

\section{Scaling of the Volume Transport $T_n$} \label{sec_scaling}
Next, let us discuss the scaling of the volume transport $T_n$ 
with the meridional pressure difference $\Delta p$. First, consider the situation without the 
SO processes, i.e. $T_s^{(e)}=T_s^{(gm)}=0$. The scaling can be 
obtained from the general solution in equ.~(\ref{equ_solution}) with 
$\gamma_e=\gamma_{gm}=0$. More illustrative is the derivation from the 
original equations for the volume transport (\ref{equ_tn}) and (\ref{equ_tu}). 
The fact that the northern downwelling has to be balanced by the low-latitude 
upwelling
$ T_n = \gamma_n D \cdot \Delta p = T_u = \gamma_u/D$
implies that 
\begin{equation}
D^{(noSO)} = \sqrt{\frac{\gamma_u}{\gamma_n \Delta p}} 
\sim \left( \Delta p \right)^{-1/2},
\label{equ_d_noso}
\end{equation}
i.e. the PD decreases with increasing pressure difference in the 
NA.
Using this expression to replace $D$ in the parameterisation of $T_n$ 
in equ.~(\ref{equ_tn}) yields
\begin{equation}
T_n^{(noSO)} = \sqrt{\gamma_u \gamma_n \cdot \Delta p} \sim \sqrt{\Delta p}
\label{equ_noso_scaling}
\end{equation}
In connection with equ.~(\ref{equ_dp}) we get the scaling 
$T_n^{(noSO)} \sim \left(\Delta \rho \right)^{1/3}$ which was derived first by 
{\it Bryan} [1987].\\ 
%\cite{bryan87}
Next let us add the SO winds, but neglect the eddy-induced return flow, i.e. $\gamma_{gm}=0$. 
The solution~(\ref{equ_solution}) then becomes
$T_n = \gamma_e + \sqrt{\gamma_e^2 + \gamma_u \gamma_n \Delta p }$,
which goes to a constant $T_n \rightarrow 2 \gamma_e = T_s^{(e)} = const.$ in the wind-driven limit, 
i.e. for small vertical diffusivity $\gamma_u \ll \gamma_e^2 / (\gamma_n \Delta p)$. 
As expected no scaling between pressure difference and volume transport is observed in this case.

\begin{figure}%[h!]
\begin{center}
\includegraphics[width=18pc]{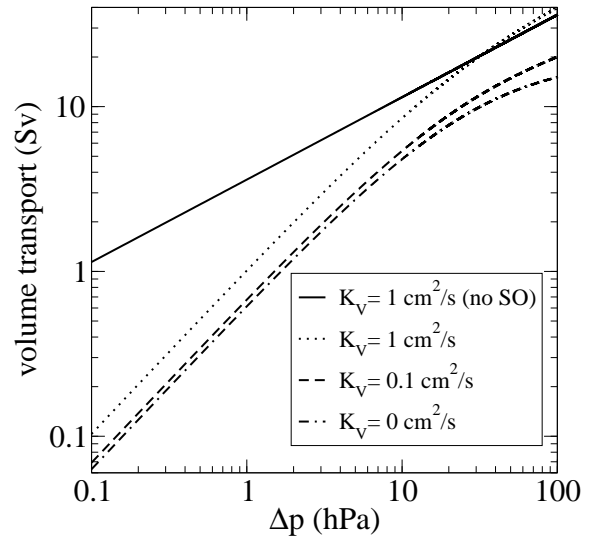}
\caption{\label{fig_analytic_solution_log} The same curves as in fig.~\ref{fig_analytic_solution} in 
double logarithmic scale. The dashed line corresponds to the solution without the 
SO processes and shows a one-half scaling $T_n \sim \sqrt{\Delta p}$. In 
contrast to this the solutions which include the SO processes show a linear
relationship $T_n \sim \Delta p$.}
\end{center}
\end{figure}

In order to discuss the scaling behaviour in the presence of both SO processes in the model, 
we plot the solutions shown in 
fig.~\ref{fig_analytic_solution} in double logarithmic scale in 
fig.~\ref{fig_analytic_solution_log}. For small 
$\Delta p$ all solutions which include the SO processes 
have slope one which corresponds to a linear scaling of the volume transport with the 
pressure difference. For comparison the solution without the SO 
processes from equ.~(\ref{equ_noso_scaling}) has been included as the solid 
curve in fig.~\ref{fig_analytic_solution_log} showing the one-half slope. 
This result can be understood from the general solution in equation (\ref{equ_solution}) 
which also sets the scale $\Delta p_s$ for which the linear relation holds. 
For $\Delta p \ll \Delta p_s \equiv \gamma_{gm}/\gamma_n$ 
the solution~(\ref{equ_solution}) becomes
\begin{equation}
T_n = \frac{\gamma_n \gamma_e}{\gamma_{gm}} 
\left(1 + \sqrt{1 + \frac{\gamma_u \gamma_{gm}}{\gamma_e}}\right) \cdot \Delta p
\sim \Delta p, 
\label{equ_small_dp}
\end{equation}
i.e. $T_n$ is linear in the meridional pressure difference $\Delta p$. Using the numerical 
values given by 
{\it G99}, we obtain an estimate for the pressure scale 
$\Delta p_s = 31.25 \; hPa$ which is consistent with the scaling seen in 
fig.~\ref{fig_analytic_solution_log}. Simulations with the oceanic general 
circulation model MOM-3 show a linear scaling of $T_n$ with $\Delta p$ for a 
variety of parameter settings, including the case of zero diapycnal mixing 
({\it Griesel}, (pers. comm., 2004). 
The pressure scale in these simulations is of the 
order of $\Delta p_s \approx 50 hPa$ which is in good agreement with the above 
estimate.

The physical meaning of the scaling regime is seen when 
multiplying $\Delta p_s$ with $D$ which gives
$T_n \ll T_s^{(gm)}$.
which means that the scaling occurs only when the circulation is completely dominated by 
the SO processes, i.e. when the eddy-induced return flow in the Southern 
Ocean is much stronger than the downwelling in the NA. This situation 
is not consistent with the underlying physical assumption of the model of an 
interhemispheric meridional overturning circulation and it does not describe the 
observed circulation in the ocean.
From equ.~(\ref{equ_D_tn_dp}) 
and (\ref{equ_small_dp}) we can see that in the linear scaling regime the pycnocline 
depth does not vary with the pressure difference, in contrast to the situation 
without SO processes (equ.~(\ref{equ_d_noso})) where $D$ decreases 
with $\Delta p$.
From equ.~(\ref{equ_dp}) we see that for constant $D$ the pressure difference
scales in the same way as the density difference $\Delta \rho \sim \Delta p$ making 
the linear scaling a simple consequence of the initial assumption that 
$T_n \propto \Delta \rho$ (equ.~\ref{equ_tn}). 

%-figure:- analytic_solution_log

\section{Comparison with simulations} 
\label{ogcm_comparision}
\begin{figure}%[h!]
\begin{center}
\includegraphics[width=18pc]{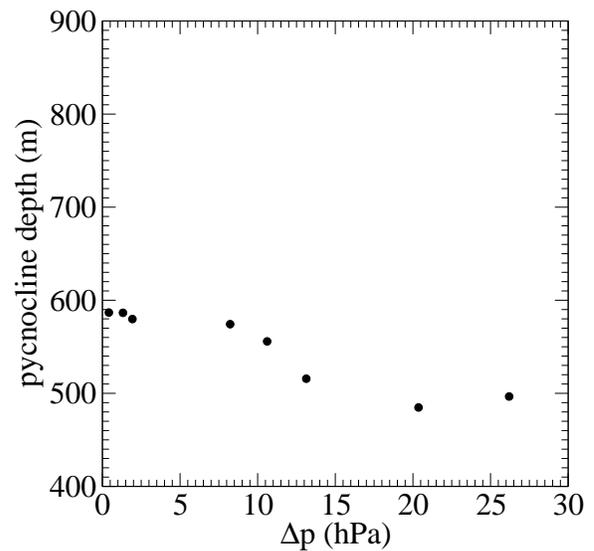}
\caption{\label{fig_pycnocline} Pycnocline depth for the simulations shown 
in fig.~\ref{fig_analytic_solution} as a function of the meridional pressure 
difference $\Delta p$ in the NA. Definition and displayed depth range were taken 
as in {\it G99}.}
\end{center}
\end{figure}

The linear relationship between the maximum overturning strength and the density 
difference $\Delta \rho$ which was observed by {\it Rahmstorf} [1996] in a 
oceanic GCM is reflected in the parameterization of the 
northern downwelling in equ.~(\ref{equ_tn}). In the conceptual model, however, 
it does not carry over to the pressure difference, as was shown in the previous section. 
In order to check this scaling we 
carried out simulations with the coupled climate model CLIMBER-3$\alpha$. The 
model contains an atmosphere and a sea-ice module as well as the oceanic general 
circulation model MOM-3. 
The effect of baroclinic eddies was included through a 
parameterization following {\it Gent and McWilliams} [1990] with a coefficient of 
$\kappa_{gm} = 2.5 \cdot 10^{6} cm^2 s^{-1} $. 
For a full description of the model see {\it Montoya et al.} [2004]. 
Starting from the present day equilibrium simulation with a maximum overturning strength of
12 Sv, we apply a negative salinity forcing of different strength to 
the NA convection sites (between $50^\circ$N and $80^\circ$N) as 
described in {\it Levermann et al.} [2004]. This 
leads to a decrease in the meridional pressure difference in the NA and therefore 
a weakening of the meridional overturning. A positive salinity forcing strengthens 
the overturning and increases the pressure difference.
Fig.~\ref{fig_analytic_solution} shows the simulations as black dots. The pressure 
was taken at a depth of 1500 m corresponding to the center of the overturning cell in the simulations. 
The differences were taken between the zonal average between $50^\circ$N and $80^\circ$N and 
the zonal average between $20^\circ$N and $30^\circ$N. This corresponds with the 
meridional pressure difference in the NA that enters 
equ.~(\ref{equ_tn}).
As seen in fig.~\ref{fig_analytic_solution} the maximum meridional overturning 
in the Atlantic scales linearly with the pressure difference in the NA 
in the simulations. The vertical diffusivity in coupled model was kept constant 
at $\kappa_v = 0.1 \; cm^2 s^{-1}$. Thus the simulations correspond to the 
dashed solution curve in fig.~\ref{fig_analytic_solution}. Simulations and conceptual model do neither 
agree quantitatively using the values suggested by {\it G99} nor is the 
qualitative behaviour of the two main quantities (pressure and volume transport) 
reproduced in the conceptual model. These results are supported by recent findings by 
{\it Griesel} (pers. comm., 2004) with an oceanic GCM. Their 
work shows that the linear scaling between pressure and overturning strength 
is a robust feature. It is independent of changes to various parameters including 
the Gent and McWilliams diffusivity coefficients.
In order to emphasize that fact that the linear scaling $T_n \sim \Delta p$ 
corresponds to constant $D$ we plot in fig.~\ref{fig_pycnocline} 
the PD as defined in {\it G99} for our simulations.
In contrast to {\it G99} in an OGCM we find in our coupled model 
no significant variation of the PD for varying pressure difference.

%-figure:- analytic_solution_log

\section{Conclusions}
By giving an analytic expression for the meridional overturning strength $T_n$ as a function 
of the meridional pressure difference $\Delta p$, we discuss the scaling of the 
two main quantities of the conceptual model introduced by {\it G99}. 
The model exhibits two scaling regimes which both correspond to unphysical situations.
Linear scaling occurs in a situation where the 
eddy-induced return flow is much stronger than the northern downwelling. This 
corresponds to a circulation which is localized entirely in the SO and 
in which all downward volume transport is due to the eddy-induced return flow. 
This situation is inconsistent with the physical assumption of an interhemispheric 
overturning cell underlying the model and the isopycnal nature of the return flow.

The second scaling regime corresponds to a purely northern cell where the upwelling 
takes place entirely in low latitudes, described by an advection-diffusion balance. 
In this case the overturning is proportional to the square root of the pressure as 
reported by {\it Bryan} [1987]. 
The scaling was checked using the coupled climate model CLIMBER-3$\alpha$ with a parameter 
setup comparable to the conceptual model, i.e. including effects of 
baroclinic eddies following {\it Gent and McWilliams} [1990] and a vertical diffusivity 
of $\kappa_h = 0.1 cm^2 s^{-1}$. The simulations exhibit a linear scaling and 
therefore support previous studies [{\it Hughes and Weaver}, 1994; {\it Rahmstorf}, 1996; 
{\it Thorpe et al.}, 2001] with comprehensive 
climate models. The PD does not vary significantly as a function 
of the pressure difference in our simulations.

We conclude that the conceptual model of the PD can not 
reproduce the scaling between its central variables, the pressure and the 
volume transport. Besides possible criticism regarding the specific parameterizations 
of the four physical processes contained in the model, the assumption of a 
universal $D$ for all these processes seems questionable. 

\acknowledgments
A.L. was funded by the Gary Comer foundation. A.G. were funded through the 
James S. McDonnell Foundation Centennial Fellowship. The authors are grateful 
to two anonoumous referees for enriching comments.

\end{article}

\end{document}